\documentclass[a4paper]{jpconf}
\usepackage{graphicx}
\begin{document}
\title{High temperature spin-triplet topological superconductivity in K$_2$Cr$_3$As$_3$} 

\author{Guo-qing Zheng}

\address{Department of Physics, Okayama University, Okayama 700-8530, Japan}

\ead{zheng@psun.phys.okayama-u.ac.jp}

\begin{abstract}
Spin-triplet superconductors are novel materials capable of harboring Majorana bound states that can be used in topological quantum computing. However,  such bulk  materials are still rare. Here we review the results that established
 K$_{2}$Cr$_{3}$As$_{3}$    as a spin-triplet superconductor with transition temperature $T_c$ as high as 6.5 K. We focus on the multiple-phases feature, and its exquisite distance   to a ferromagnetic quantum critical point which is likely responsible for the high $T_c$. We  touch on the topological aspect of the superconducting state, and suggest that it is 
a new route to the technical implementation using a topological spin-triplet superconductor at the highest temperature ever.
\end{abstract}

\section{Introduction}

In contrast to a spin-singlet state in materials including  copper-oxide and iron-pnictide high temperature superconductors, where the electron pairs (Cooper pairs) are in the  state with total spin $S$=0 \cite{cuprates,MatanoEPL},   
spin-triplet superconductivity or fluidity  with $S$=1 possesses internal structure, 
which gives rise to rich  physics such as additional symmetry breaking,  order parameter collective modes, 
and multiple phases of the condensate. 
Therefore, spin-triplet superconductors are an excellent playgound for testing and creating new concepts of condensed matter physics. In addition, a spin-triplet state with odd-parity  can be topological and  
Majorana bound states or chiral Majorana fermions can emerge in the vortex cores or   on boundary, which are robust against perturbation. Thus, spin-triplet superconductivity are  of great  interests and importance not only in fundamental physics, but also  in applications as their edge or bound states  can be used to implement topological quantum computing based on  non-Abelian 
statistics~\cite{Kitaev}.
However, such materials are still rare. 
Historically, efforts of searching for spin-triplet superconductivity was devoted to strongly correlated electron systems. 
In  recent years, doped topological insulators have been proposed to be a new playground for spin-triplet superconductivity, and indeed unambiguous evidence has been obtained for   Cu$_x$Bi$_2$Se$_3$ through the discovery of spin-rotation symmetry breaking by nuclear magnetic resonance (NMR) \cite{Matano}.

More recently, a new superconducting  family containing 3d transition-metal element  Cr,  A$_{2}$Cr$_{3}$As$_{3}$ (A = Na, K, Rb, Cs) was discovered~\cite{K, Rb, Cs, Na}, with  $T_{\rm c}$ as high as 8 K. Early NMR measurements in a poly-crystal sample found  ferromagnetic  spin fluctuation in the normal state and point nodes in the gap function in the superconducting state \cite{YangPRL}, indicating that this family is a strongly-correlated 3d electron system.
Recent Knight shift measurements in a single crystal of K$_{2}$Cr$_{3}$As$_{3}$  with  $T_c$ = 6.5 K revealed that the spin susceptibility   is unchanged below $T_{\rm c}$ when the magnetic field $H_{\rm 0}$ is applied in the  $ab$ plane, but vanishes toward zero  when $H_{\rm 0}$ is along the $c$ axis, which unambiguously establishes this compound as a spin-triplet superconductor described by a vector order-parameter   $\textbf{{d}}$
parallel to the $c$-axis at low fields \cite{YangSciAdv}. 
In this presentation, we focus on  the multiple phases 
of K$_2$Cr$_3$As$_3$ and its equisite clossness to a ferromagnetic quantum critical point (QCP), which may be responsible for the high $T_c$. We will also touch on the topological aspects of the superconducting state.

\section{Multiple superconducting phases and rotation of the vector order parameter at high fields}

\begin{figure}[htbp]
	\includegraphics[clip,width=90mm]{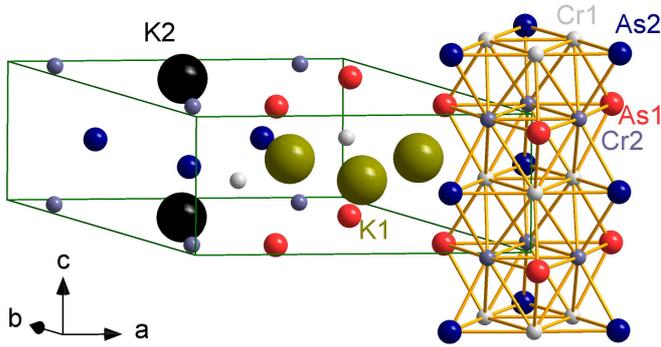}
	\caption{\label{crystalstructure}
	Crystal structure of K$_2$Cr$_3$As$_3$. The Cr$_3$ tube runs along the $c$-axis,  surrounding which runs the As$_3$ tube. 
	
	}
\end{figure}

Figure 1 shows the crystal structure of K$_2$Cr$_3$As$_3$. There are two Cr sites, both of which form a triangle shape. The Cr$_3$ tube runs along the $c$-axis, which is the reason why the compound is often called 
"quasi-one dimensional". However, 
band calculation shows that there are three pieces of Fermi surfaces, among which  the three-dimensional one makes the main contribution to the density of states (DOS)  \cite{Jiang}. Also, very recent neutron scattering measurements found that spin fluctuations due to nesting of the two one-dimensional Fermi surfaces do not contribute to superconductivity that must exist in the three-dimensional Fermi surface \cite{NS}

\begin{figure}[htbp]
	\includegraphics[width=17cm]{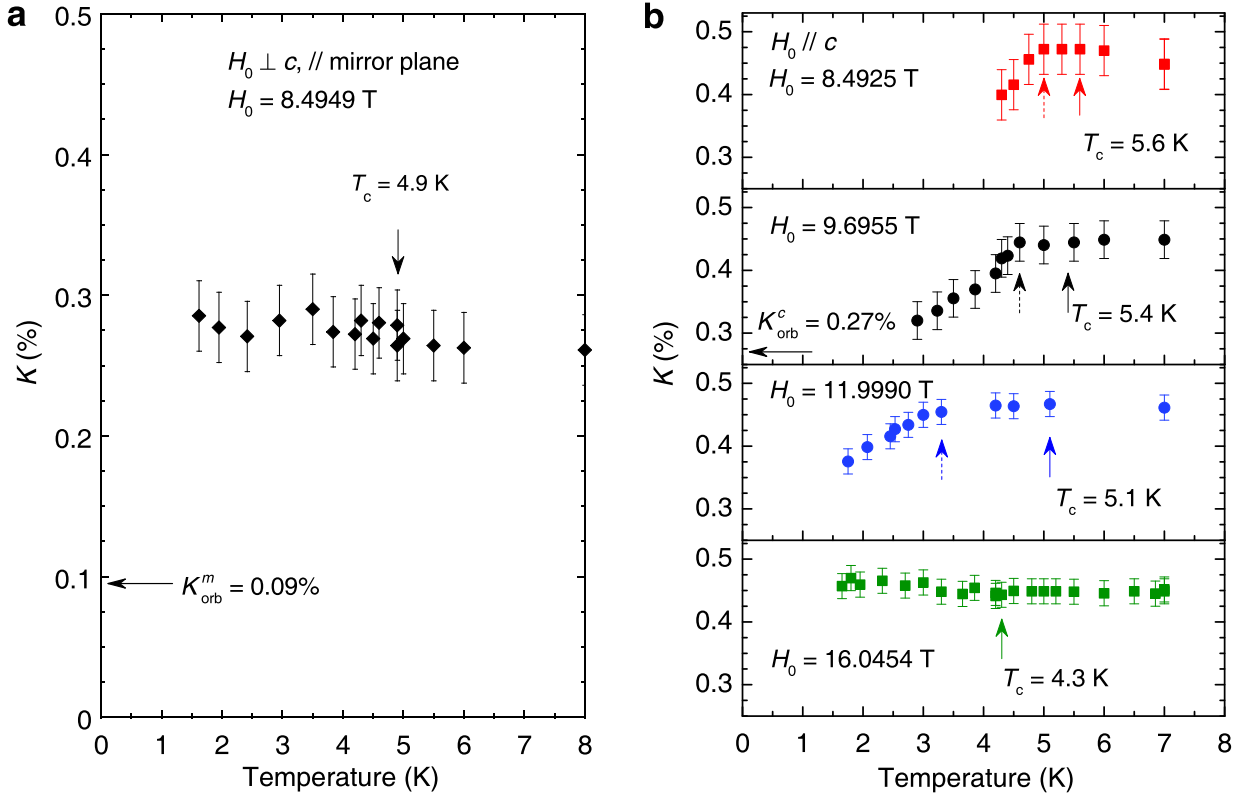}
	\caption{
		(
		(a) The temperature dependence of the Knight shift with the magnetic
		field applied in the $ab$ plane (and along the mirror plane of the lattice).
		(b) The Knight shift with  $H_{\rm 0} \parallel$ $c$ axis. The solid arrows indicate $T_{\rm c}$ determined by ac susceptibility, spin-lattice relaxation rate $1/T_1$ and the NMR intensity. The dotted arrows indicate the temperature $T^{*}$ below which the Knight shift starts to drop.
		\label{KSC}}
\end{figure}

Figure 2 shows the $^{75}$As-NMR Knight shift in a single crystal above and below $T_c$ for different field orientations. The Knight shift $K$ consists of two contributions, namely, $K_s$ from spin susceptibility and $K_{\rm orb}$ from orbital susceptibility.
The diamagnetic contribution due to a vortex lattice formation in the superconducting state is negligibe as the penetration depth is long \cite{YangSciAdv}. The position of $K_{\rm orb}$ is shown by the horizontal arrows in Fig. 2. It was estimated from a novel analysis using the relation between $K$ and the spin-lattice relaxation rate devided by $T$, $1/T_1T$. Recently, by using a more conventional method, the so-called $K- vs -\chi$ plot,  very close values have been obtained \cite{Ogawa}. Thus, $K_s$ has a substantial contribution for both $H\parallel c$ and  $H\perp c$
directions.

As can be seen  in Fig. 2, $K$ (and thus $K_s$) shows a very anisotropic temperature dependence below $T_c$ between $H_0\parallel c$ and  $H_0\perp c$. Namely, For  $H_0\perp c$, $K_s$ does not change across $T_c$. In contrast, for $H_0\parallel c$, $K_s$ is reduced in the superconducting state. In particular, For $H_0$=9.6955 T, $K_s$ decreases toward zero in the $T$=0 limit. These results are firm evidence of spin-rotation symmetry breaking or an emergence of spin nematicity, and   establish  K$_2$Cr$_3$As$_3$ as a spin triplet superconductor described by the vector order parameter $\textbf{{d}} \parallel c$-axis. 

A novel aspect of Fig. 2 is that  $K_s$ does not decrease right at $T_{\rm c}$ but at a lower temperature $T^*$. 
It is emphasized that both the NMR intensity and $1/T_{\rm 1}$ drop sharply at $T_{\rm c}$.
The disparity between   $T_{\rm c}$ and  $T^*$ becomes larger as $H_0$ increases. Most intriguingly, under $H_0$=16.0454 T at which $T_c$($H$)=4.3 K,  $K$ shows no reduction at all.  
These results are summarized in Fig. 3.

\begin{figure}[htbp]
	\includegraphics[width=8.5cm]{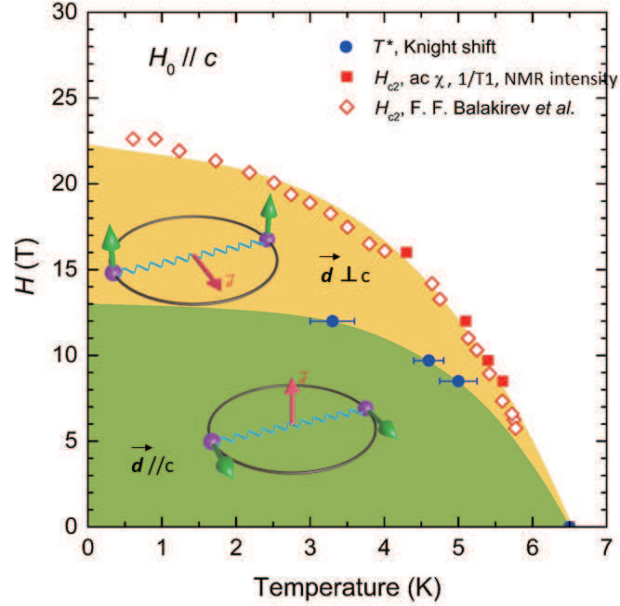}
	\centering
	\caption{$H$-$T$ phase diagram for K$_{2}$Cr$_{3}$As$_{3}$ with the field applied along the $c$-axis. The shaded regions are the superconducting state. In the illustration, the purple balls represent electrons that form a Cooper pair, the wavy lines depict the attractive force between two electrons. The green trigonal (pyramid) heads indicate the spin direction and the red arrorws show the direction of the vector order parameter $\textbf{{d}}$. 
		The Pauli-limit field is estimated to be 13 T at $T$=0, being consistent with the boundary between the green and yellow regions.
		The data points marked by open diamonds are from Ref. \cite{Canfield}.
		\label{PD-2}}
\end{figure}

The boundary marked by $T^*$ in Fig.3 is consistent with the Pauli-limited field $H_p$ which is estimated to be about 13 T at $T$=0. It is thus consistent with $\textbf{{d}} \parallel c$-axis for the low fields as discussed abouve.   Above the $T^*$  boundary, as the Knight shift shows no reduction, the  $\textbf{{d}}$-vector must be flipped by 90 degrees. The mechanism for the $\textbf{{d}}$-vector rotation is unclear at the moment. There are two possibilities. One is the SO(5)-like transition to gain Zeeman energy. The other is that the  $T^*$  boundary marks a phase transition from one symmetry to another. Measurements at high fields are now in progress to resolve this issue. 
In any case, the multiple-phases feature is a direct consequence of a spin-triplet condensate due to the degrees of  internal freedom of the pairing and provides good opportunity to investigate novel phases of matter. 

\section{Why $T_c$ is high?}

\begin{figure}[hbp]
	\includegraphics[width= 9.5 cm]{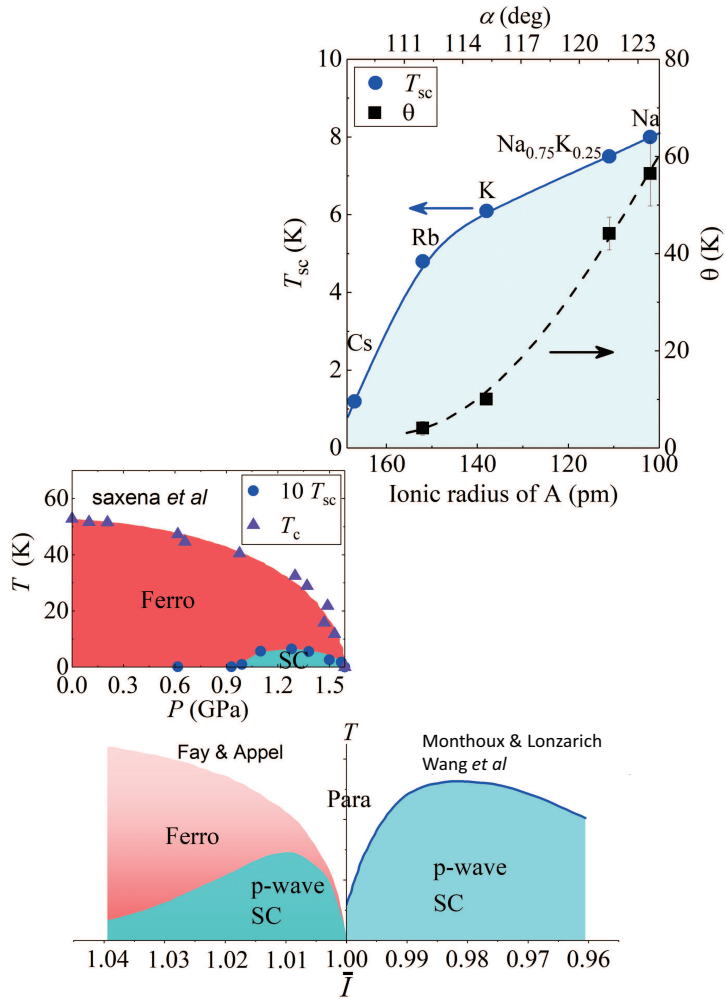}
	\caption{ (a) Phase diagram of the A$_{2}$Cr$_{3}$As$_{3}$ (A = Na, Na$_{0.75}$K$_{0.25}$, K, Rb, Cs) family.  The $\alpha$ is the angle of the Cr2-As2-Cr2 bonds (see Fig. 1). In this figure, $T_{\rm C}$ is reserved to represent the Curie temperature and the superconducting transition temperature is marked by $T_{\rm sc}$.
		(b) Phase diagram of UGe$_2$ under pressure.  \cite{UGe2}. 
	  (c) Theoretically proposed phase diagram for superconductivity in ferromagnetic  ("Ferro") ordered state \cite{appel} and paramagnetic ("Para") state with ferromagnetic correlation.
			\cite{Monthoux,WangZQ}. 
		The horizontal axis is the exchange interaction $I$ with $I$ = 1 at the QCP.
		\label{poly-PD}}
\end{figure}

The Alkali element A in the  formula A$_{2}$Cr$_{3}$As$_{3}$ can be changed and superconductivity with a maximal $T_c$ =8 K was obtained. A natural question is what controls $T_c$ within the family and why  $T_c$ is much higher than other systems with ferromagntic interactions. 
To answer the question, the temperature dependence of  $1/T_1T$ was measured across the family and  analyzed \cite{LuoPRL}. The quantity
1/$T_1T$ measures spin correlations through probing the imaginary part of transverse dynamic susceptibility $\chi$''(\textbf{$\omega$,q}) summed over all \textbf{q}. 
In the case of strong electron correlations,  we may write 1/$T_1T$  as: 
\begin{equation}
1/T_1T = (1/T_1T)_{\rm SF} + (1/T_1T)_{0}. 
\end{equation}
where the first part  originates from spin fluctuations and the  second part is due to the DOS at the Fermi level. According to self-consistent renormalization theory for three-dimensional ferromagnetic fluctuation \cite{Moriya},  (1/$T_1T$)$_{\rm SF}$ follows a Curie-Weiss temperature dependence 
\begin{equation}
(1/T_1T)_{\rm SF} = C/(T + \theta) 
\end{equation}
Here, $\theta$ describes the distance to a ferromagnetic QCP. The obtained   parameter $\theta$ from the fittings is shown in  Fig.~\ref{poly-PD}(a). The  $\theta$ gradually increases from 4.1 K for Rb$_2$Cr$_3$As$_3$ to 56.8 K for Na$_2$Cr$_3$As$_3$. 

 Figure~\ref{poly-PD} compares our phase diagram of the A$_2$Cr$_3$As$_3$ family with  theories for ferromagnetic superconductors \cite{appel,Monthoux,WangZQ} and that  of UGe$_2$\cite{UGe2}. In this figure, the superconducting transition temperature is represented by $T_{\rm sc}$, in order to distinguish between the Curie temperature  $T_{\rm C}$.  In Fig.\ref{poly-PD}(a) is plotted the parameter $\theta$ and $T_{\rm sc}$  as a function of the ionic radius of the alkali element A. There is a strong correlation between the ionic radius and the  Cr2-As2-Cr2 bond  angle $\alpha$ (see Fig. 1) which is shown as the upper horizontal axis.   
The angle $\alpha$ may be related to double exchange interaction through Cr-As-Cr path.
 At  $\alpha$=90$^\circ$, the As-4$p_x$ and As-4$p_y$ orbitals become degenerated with respect to Cr-3$d$ 
 orbitals,
 which will  maximize the double exchange between 
 the two Cr2  along the $c$-axis via the As-4$p_x$ and As-4$p_y$ orbitals \cite{magnetism}.
 Therefore, on going from A = Na to Na$_{0.75}$K$_{0.25}$, K, and  Rb, 
 an increase in the  ferromagnetic interaction can be expected,  
 which drives the system towards 
 a ferromagnetic  QCP. Our finding that Na$_2$Cr$_3$As$_3$ has a weaker spin correlation was subsequently supported by a theoretical evaluation based on the first-principle calculation \cite{ChenQJ}
 
 In Fig.~\ref{poly-PD}(a), 
 one sees that  \emph{T}$_{\rm sc}$ increases  upon moving away from the putative  ferromagnetic QCP. This is in sharp contrast to the antiferromagnetic case where \emph{T}$_{\rm sc}$ usually forms a peak around a QCP, but is consistent with theoretical expectations.  
 Superconductivity  in both the ferromagnetic ordered state and near a ferromagnetic QCP  in paramagnetic side was first discussed by Fay and Appel \cite{appel}.  Later on, Monthoux and Lonzarich \cite{Monthoux}, and  Wang {\it et al} \cite{WangZQ} improved the calculation for the paramagnetic side. 
 In the antiferromagnetic case, the pairing interaction is enhanced due to increased quantum fluctuations \cite{Monthoux}.
 In the ferromagnetic case, when it is approached from the paramagnetic side,  increased quantum fluctuations also enhances pairing strength \cite{appel,Monthoux}, 
 but
 mass enhancement and a finite quasiparticle life time  act as  pair breaking and suppress \emph{T}$_{\rm sc}$ \cite{appel,Monthoux,WangZQ}. 
  As can be seen in Fig.~\ref{poly-PD}(a), K$_{2}$Cr$_{3}$As$_{3}$ is away from the ferromagnetic QCP yet with spin correlation, which probably explains why \emph{T}$_{\rm sc}$ is high.

\section{Topological aspects}	

Having established the spin symmetry, the next issue is the orbital wave function of the Cooper pairs. Group theory analysis \cite{YangSciAdv} shows that only  $E'$ states ($p_x$+$ip_y$ and $p_x$-$ip_y$) are
consistent with the Knight shift result 
and the $1/T_1$ result revealing a nodal gap \cite{YangPRL,LuoPRL}. This state is analogous to the A phase (or Anderson-Brinkman-Morel state) 
in superfluid $^{3}$He, and 
 is topological. Therefore 
Majorana zero modes can be expected in vortex cores \cite{Ivanov}. 
In particular, if a superconducting thin film with its thickness  smaller than the superconducting coherence length is available,  a single Majorana zero mode will be expected in the core of a half-quantum vortex.
Table I lists the topological classification together with Cu$_x$Bi$_2$Se$_3$ ($x<$  0.46) \cite{Matano}, Cu$_x$Bi$_2$Se$_3$ ($x\geq$0.46) \cite{Kawai} and Li$_2$Pt$_3$B \cite{Nishiyama}.

\begin{table}[h]
	\centering
	\caption{Topological aspects of K$_{2}$Cr$_{3}$As$_{3}$ and other superconductors and superfluid $^3$He.}
	
	\begin{tabular}{|c|c|c|c|}
		\hline\hline
		Class &Time Reversal Symmetry&Two Dimension &Three Dimension\\ 
		\hline
		D &Broken &$^3$He-A film &  \\
		
		&       & K$_{2}$Cr$_{3}$As$_{3}$ film & \\
		
		&       &  Cu$_{x}$Bi$_{2}$Se$_{3}$ ($x\geq$ 0.46) film & \\
		\hline
		
		
		DIII     & Conserved & & $^3$He-B \\
		&  & & Li$_2$Pt$_3$B  \\
		&  & & Cu$_{x}$Bi$_{2}$Se$_{3}$ ($x<$ 0.46) \\

		\hline\hline
	\end{tabular}	
\end{table}

\section{Summary}

By NMR measurements in a single crystal of  K$_{2}$Cr$_{3}$As$_{3}$, we discover a  nematic response of the spin susceptibility between $H_0\parallel c$-axis and $H_0\perp c$-axis in the superconducting state, establishing this compound as a spin-triplet superconductor described by the vector order parameter $\textbf{{d}}$ parallel to the $c$-axis. However, at high fields above 13 T, the $\textbf{{d}}$-vector is flipped to lie in the $ab$-plane. The multiple phases at different magnetic fields is another manifestation of the spin-triplet pairing due to its internal degrees of freedom and merits further investigation.
Our results demonstrate that K$_{2}$Cr$_{3}$As$_{3}$  is a new platform for basic research of topological materials  and technical applications of topological superconductivity.

\section*{Acknowledgments}
	This work was performed in collaboration with J. Yang, J. Luo, Y. Zhou,  Y.G. Shi, Z.A. Ren, G.H. Cao, S. Ogawa, K. Matano, and Y. Inada. We thank T. Hanaguri and K. Machida for stimulating discussions.
	The work was partially supported by  JSPS Grants ( JP19H00657 and JP22H04482).

\end{document}